\documentclass[]{aastex631}

\usepackage{amsmath}

\def\rb{r_b}
\def\vb{v_j}
\def\nel{n_{{\rm e}^-}}
\def\np{n_{\rm p}}
\def\mel{m_{{\rm e}}}
\def\mp{m_{\rm p}}
\def\rhoj{\rho_j}
\def\thetaj{\Theta_ j}

\def\rhoa{\rho_ a}
\def\thetaa{\Theta_a}

\usepackage{comment}
\graphicspath{{./}{}}
\begin{document}

\title{The Morphology and Dynamics of Relativistic Jets with Relativistic Equation of State}

\author[0000-0002-9036-681X]{Raj Kishor Joshi}
\affiliation{Aryabhatta Research Institute of Observational Sciences (ARIES) \\
Manora Peak  \\
Nainital 263001, India  }

\affiliation{Department of Physics, Deen Dayal Upadhyay Gorakhpur University \\
Gorakhpur, 273009, India  \\
}

\author[0000-0002-2133-9324]{Indranil Chattopadhyay}
\affiliation{Aryabhatta Research Institute of Observational Sciences (ARIES) \\
Manora Peak  \\
Nainital 263001, India  }

\begin{abstract}
We study the effect of plasma composition on the dynamics and morphology of the relativistic astrophysical jets. Our work is based on a relativistic total variation diminishing (TVD) simulation code. We use a relativistic equation of state in the simulation code which accounts for the thermodynamics of a multispecies plasma which is a mixture of electrons, positrons, and protons.  
To study the effect of plasma composition we consider various jet models. These models are characterized by the same injection parameters, same jet kinetic luminosity, and the same Mach numbers. The evolution of these models shows that the plasma composition affects the jet head propagation speed, the structure of the jet head, and the morphology despite fixing the initial parameters. We conclude that the electron-positron jets are the slowest and show more pronounced turbulent structures in comparison to other plasma compositions. The area and locations of the hot-spots also depend on the composition of jet plasma. Our results also show that boosting mechanisms are also an important aspect of multi-dimensional simulations which are also influenced by the change in composition. 
\end{abstract}


\section{Introduction} \label{sec:intro}
Collimated relativistic outflows or jets are a common feature of active galactic nuclei (AGN) as well as of microquasars (i. e., an X-ray binary where a stellar-mass black hole/neutron star is the primary and a normal star is the secondary). These jets carry a large amount of kinetic energy which is transported away from the center of gravity. Many of these jets are radio bright. Depending upon the energy, the radio emitting extragalactic jets (i. e., from AGN) are classified into two categories namely FR-I and FR-II \citep{fr74}. The FR I are the low luminosity jets with brighter structures close to the nucleus and the FR II sources have powerful jets and the maximum brightness is at the regions of jet termination. Apart from the difference in power, possibly the interaction between the jet beam and external medium, the effect of entrainment, and mixing due to different types of jet instabilities \citep{rbmc20} also play a significant role in shaping the FR-I and FR-II morphology. The understanding of the morphology and dynamics of the jets has significantly improved by analytical \citep{br74,k94,kk06,bnps11} and numerical studies \citep{mmi97,mhd97,myt04,shr21}. All these numerical simulations show that the jets produce complex structures with shocks, back flow, and instabilities when they interact with the ambient medium. The growth in computational power and improvement in numerical algorithms has given a tremendous boost to this subject. The high-resolution shock capturing (HRSC) simulation methods \citep{mmi94,dh94,wmkm13} improved the understanding of the morphology of relativistic jets. Recently, the numerical studies have been expanded to include the effect of magnetic field on jet dynamics through magnetohydrodynamic (MHD) simulations \citep{kmvc08,mrbfm10,hk14,kp21}. Numerical simulation codes solve the set of conservation laws along with an equation of state (EoS) which relates the thermodynamic variables like internal energy, pressure, and mass density. Most of the simulation codes are based on the ideal gas equation of state, in which the adiabatic index remains a constant parameter (4/3 for hot gas and 5/3 for cold gas). This EoS is a reasonable approximation if the flow remains ultra-relativistic or non-relativistic throughout its evolution. But for the astrophysical jets which span over very large length scales and vary over a large span of temperatures and propagation velocities, the use of fixed adiabatic index EoS is not physically consistent. Jets also exhibit strong shocks where the gas is heated to extreme temperatures. In the context of astrophysical jets, \cite{dh96} highlighted the importance of using an EoS which self-consistently calculates the adiabatic index. The EoS for relativistic fluid has been computed independently by various authors \citep{c39,s57}. However, the expressions for these EoS involve complicated modified Bessel functions, hence, their implementation in the simulation codes is a complicated task and it increases the computational cost as well \citep{fk96}. To avoid this extra computational cost and numerical difficulties, many algebraic approximations have been proposed \citep{m71,mpb05,rcc06} for the relativistic EoS. These EoS self-consistently calculate the adiabatic index from the information of temperature and account for the transition between cold to hot thermal states (and vice versa) in a consistent manner. It was shown that the EoS proposed by \cite{rcc06} mimics the one by \citeauthor{c39}, extremely well.
The EoS proposed by \citep{rcc06} was extended by \citet[][abbreviated as CR EoS]{cr09} for the plasma composed of dissimilar particles. CR EoS can account for a mixture of electrons, positrons, and protons which allows us to study the effect of plasma composition on the dynamics of astrophysical flows. CR EoS has been used in many analytical investigations and the effect of plasma composition was studied  \citep{cc11,ck16,sc19,scp20,jcy22,sc22}. However, to date, there is a lack of numerical investigations to study the effect of plasma composition on jet dynamics \citep[with a notable exception of][]{samgm02}. In our previous analytical investigation \citep{jcry21}, we obtained the exact solution of one-dimensional jet using CR EoS. We investigated when a Riemann problem will behave like a jet (forward shock or FS - Contact discontinuity or CD - reverse shock or RS) and when it will behave like a shock tube test (FS - CD - rarefaction fan or RF). We also studied
the effect of plasma composition on the propagation speed of the jet head and on the shock dynamics. Based on the algorithms of \cite{rcc06, crj13} we developed a one-dimensional relativistic code that incorporates gravity in the weak field limit to study radiatively driven, transonic, relativistic jets around black holes \citep{jdc22}. In this paper, we extend the work of \cite{jcry21} to two-dimensional axis-symmetric relativistic jet simulations. The metric is special relativistic but in cylindrical geometry in the space dimension. Since in multi-dimension the injected jet material is deflected back and interacts with the injected jet material the evolution of morphology is likely to be quite different from the one-dimensional study. We do see in multi-dimensional simulations, the propagation speed of the jet-head (CD) can be higher or lower than the one-dimensional estimate depending on how many RFs are produced. We would like to study it while using the CR EoS. Moreover, is the creation of RFs also affected by jet composition, which therefore affects the jet propagation speed? In this paper, we study jet dynamics in general and in particular, the effect of plasma composition on jet physics.  
The paper is organized as follows. In Section \ref{sec:setup} we describe the governing equations, details of the simulation setup, and CR EoS. In Section \ref{subsec:morph} we discuss the results describing the morphology and dynamics of the jet. Section \ref{sec:comp_effect} is devoted to show the effect of plasma composition on the jet morphology.  A brief discussion and conclusion of the work follow in Section \ref{sec:disc}.                

\section{Numerical Setup and Governing equations} \label{sec:setup}
We solve the relativistic hydrodynamic equations in two-dimensional cylindrical geometry $(r,z)$ and the three velocity vector is given by $\mathbf{v}\equiv (v^r,0,v^z)$. We use a uniform spacing grid $\Delta r=\Delta z={\rm constant}$ to discretize the computational domain in $750\times 6000$ cells. In code units, this domain is of the size $r=5,\,z=40$. The reflection boundary condition in imposed along the z-axis. The initial jet beam is resolved by $10\times 10$ cells, so the computational domain covers a region of $75\rb\times 600\rb$. The jet material is continuously injected using a fixed jet base. The outer $r$ and $z$ boundaries are kept as outflow boundaries. Assuming a beam size $\rb=0.4\,{\rm kpc}$ the computational domain covers $30\,{\rm kpc}\times 240\,{\rm kpc}$. The unit time in code is equal to $2\times 10^4$ years. The injected jet material travels through a dense, static ambient medium, and the density of the ambient medium is kept constant. We have performed simulations for various jet parameters which are listed in Table \ref{tab:model_param}. The injection velocity $\vb=0.995$ which corresponds to injection beam Lorentz factor $\gamma_j= 10$, is kept same for all the models. The time evolution of jet material is governed by the equations

\begin{subequations}
\begin{equation}
\frac{\partial D}{\partial t}+\frac{1}{r}\frac{\partial}{\partial r} \left[rDv^r\right]+\frac{\partial}{\partial z}\left[Dv^z\right]=0    
\label{eq:continuity}
\end{equation}

\begin{equation}
\frac{\partial M^r}{\partial t}+\frac{1}{r}\frac{\partial}{\partial r} r\left[M^rv^r+p\right]+\frac{\partial}{\partial z}\left[M^rv^z\right]=\frac{p}{r}    
\label{eq:momentum_r}
\end{equation}

\begin{equation}
\frac{\partial M^z}{\partial t}+\frac{1}{r}\frac{\partial}{\partial r}r\left[M^zv^r\right]+\frac{\partial}{\partial z}\left[M^zv^z+p\right]=0    
\label{eq:momentum_z}
\end{equation}

\begin{equation}
\frac{\partial E}{\partial t}+\frac{1}{r}\frac{\partial}{\partial r} r\left[(E+p)v^r\right]+\frac{\partial}{\partial z}\left[(E+p)v^z\right]=0    
\label{eq:energy}
\end{equation}
\end{subequations}

where $D,\,M^r,\,M^z,\,E$ are the conserved quantities, namely the mass density, radial and axial component of momentum density and total energy density respectively.  All these quantities are measured in laboratory frame and these are related with primitive variables as 
\begin{subequations}
\begin{eqnarray}
D=\rho\gamma \\
M^{r,z}=\gamma^2\rho h v^{r,z} \\
E=\gamma^2\rho h -p
\end{eqnarray}
\end{subequations}

 The variables $\rho,\,p,\,h$ denote proper rest mass density, pressure and specific enthalpy respectively. $v^r,v^z$ are the radial and axial components of the velocity which are related with Lorentz factor as 
 
\begin{equation}
\gamma=\frac{1}{\sqrt{1-v^2}}   \qquad {\rm with}\,\, v^2=(v^r)^2+(v^z)^2
\end{equation} 

In order to solve the set of equations \ref{eq:continuity}-\ref{eq:energy}, we need an additional equation as a closure relation. This closure equation is commonly known as equation of state (EoS).

\subsection{Equation of state}
The equation of state i.e., EoS, relates the thermodynamic variables internal energy density, rest mass density, and pressure. Most of the analytical and numerical studies use an ideal gas with a fixed adiabatic index. Fixing the adiabatic index can be a reasonable approximation only if the flow remains  either ultra-relativistic or non-relativistic. However, in the case of astrophysical jets, the use of ideal gas EoS will be inconsistent as the jets cover a very long distance and temperature varies significantly. Also, the relativistic jets are known to have multiple shock heated regions, which will contain thermally relativistic gas. Hence, it is desirable to use an EoS which accounts for self consistent evolution of the adiabatic index \citep{dh96}. We use CR EoS \citep{cr09} for relativistic multispecies fluid, which is a very close fit to the exact EoS derived by \cite{c39}. The EoS is in the form

\begin{equation}
e=\rho f
\label{eq:eos}
\end{equation}

where,
\begin{equation}
f=1+(2-\xi)\Theta\left[\frac{9\Theta+6/\tau}{6\Theta+8/\tau}\right]+\xi\Theta\left[\frac{9\Theta+6/\eta\tau}{6\Theta+8/\eta\tau}\right]
\label{eq:eos2}
\end{equation}
The mass density of fluid given as $\rho=\Sigma_i n_im_i=\nel \mel (2-\xi+\xi/\eta)$, where $\xi=\np/\nel$, $\eta=\mel/\mp$ and
$\nel$, $\np$, $\mel$ and $\mp$ are the electron number
density, the proton number density, the electron rest mass, and proton rest mass. $\Theta=p/\rho$ is a measure of temperature and $\tau=2-\xi+\xi/\eta$. The expressions for sound speed and polytropic index are given as

\begin{equation}
a^2=\frac{1}{h}\frac{\partial p}{\partial \rho}=-\frac{\rho}{Nh}\frac{\partial h}{\partial \rho}=\frac{\Gamma\Theta}{h}
\label{eq:soundsp}
\end{equation}

\begin{eqnarray}
N =  \rho \frac{\partial h}{\partial p}-1=\frac{\partial f}{\partial \Theta}
 =6\left[(2-\xi)\frac{9\Theta^2+24\Theta/\tau+8/\tau^2}{(6\Theta+8/\tau)^2}\right]+6\xi \left[\frac{9\Theta^2+24\Theta/(\eta \tau)+8/(\eta \tau)^2}{\{6\Theta+8/(\tau \eta)\}^2}\right]
\label{eq:poly}
\end{eqnarray}

The adiabatic index is related with the polytropic index as $\Gamma=1+\frac{1}{N}$, which will be a function of $\Theta$. \\

The set of relativistic hyperbolic conservation laws (\ref{eq:continuity}-\ref{eq:energy}) is solved using a relativistic total variation diminishing (TVD) routine. Originally, the TVD scheme was proposed to solve the set of non-relativistic hydrodynamic conservation laws \citep{h83}. The relativistic TVD simulation code to incorporate the relativistic EoS has been built before \citep{rcc06,crj13}. A detailed description to build the code for a general EoS is given in \cite{rcc06}. The one-dimensional relativistic TVD code with CR EoS has also been recently used to study the radiatively driven relativistic jets \citep{jdc22}.   

\section{Results}
The jet is characterized as a narrow beam of hot and relativistic material traveling through the cold and denser ambient medium. In Model OD, we present the basic structure of a relativistic jet plying through the ambient medium. The initial surface of discontinuity between the jet material and the ambient medium is the CD or the working surface. The supersonic jet material drives a shock (FS) in the ambient medium. The jet beam also contains a shock (RS), behind the contact discontinuity where the bulk kinetic energy of the jet beam is converted into  thermal energy. As the jet advances and sweeps up the ambient medium in front and drives the FS, a back flow starts to develop around the jet beam. This back flow is an important feature of highly supersonic jets. In this paper, we have studied three basic models: Models A1-A4 are jet models with the same injection parameters but we compare solutions with different compositions; Models B1-B4 are jets with the same enthalpy and jet power; Models C1-C4 are jets with the same Mach number and their initial parameters are tabulated in Table \ref{tab:model_param}.
\subsection{Model OD: The large scale morphology of the jet}
\label{subsec:morph}

\begin{table}[]
    \centering
    \begin{tabular}{c|c|c|c|c|c|c|c|c|c}
       Model & $\rhoj$ & $\thetaj$ &  $\mathcal{M}$ &  $h_j$ & $\xi$ &$L_j\,({\rm erg\,s^{-1}})$ & $\rhoa$  & $\thetaa$ & $h_a$\\
       OD  &     1.0 &     0.02   &       59.66         &   1.06& 1.00&$1.02\times10^{45}$&$2\times10^3$&$1\times10^{-5}$&1.00002\\
       A1  &     1.0 &     0.20   &      20.69         &   1.57& 0.00&$3.04\times10^{45}$&$1\times10^3$&$1\times10^{-4}$&1.000250\\       
       A2  &     1.0 &     0.20   &      23.42         &   1.76& 0.20&$3.41\times10^{45}$&$1\times10^3$&$1\times10^{-4}$&1.000253\\
       A3  &     1.0 &     0.20   &      22.89         &   1.73& 0.50&$3.35\times10^{45}$&$1\times10^3$&$1\times10^{-4}$&1.000257\\
       A4  &     1.0 &     0.20   &      22.05         &   1.67& 1.00&$3.23\times10^{45}$&$1\times10^3$&$1\times10^{-4}$&1.000259\\
       B1  &     1.0 &     0.49   &      16.33         &   2.54& 0.00&$4.92\times10^{45}$&$1\times10^3$&$5.62\times10^{-4}$&1.0014\\
       B2  &     1.0 &     0.40   &      19.13         &   2.54& 0.20&$4.92\times10^{45}$&$1\times10^3$&$5.27\times10^{-4}$&1.0014\\       
       B3  &     1.0 &     0.42   &      18.41         &   2.54& 0.50&$4.92\times10^{45}$&$1\times10^3$&$5.04\times10^{-4}$&1.0014\\
       B4  &     1.0 &     0.45   &      17.47         &   2.54& 1.00&$4.92\times10^{45}$&$1\times10^3$&$5.01\times10^{-4}$&1.0014\\
       C1  &     1.0 &     0.20   &      21.0         &   1.57& 0.00&$3.03\times10^{45}$&$1\times10^3$&$1.00\times10^{-4}$&1.000250\\
       C2  &     1.0 &     0.29   &      21.0         &   2.15& 0.20&$4.16\times10^{45}$&$1\times10^3$&$1.02\times10^{-4}$&1.000258\\       
       C3  &     1.0 &     0.27   &      21.0         &   2.00& 0.50&$3.87\times10^{45}$&$1\times10^3$&$1.03\times10^{-4}$&1.000265\\
       C4  &     1.0 &     0.24   &      21.0         &   1.82& 1.00&3.52$\times10^{45}$&$1\times10^3$&$1.04\times10^{-4}$&1.000269\\

    \end{tabular}
    \caption{Details of jet and ambient medium parameters used in different jet simulations. The variables $\rhoj\,(\rhoa),\,\thetaj\,(\thetaa),\,h_j\,(h_a),\mathcal{M},\,L_j$ represent the jet (ambient) mass density, jet (ambient) dimensionless temperature, jet (ambient) specific enthalpy, relativistic Mach number of jet beam, and the kinetic luminosity of the jet.}
    \label{tab:model_param}
\end{table}

\begin{figure}
\includegraphics[width=18cm,height=8cm]{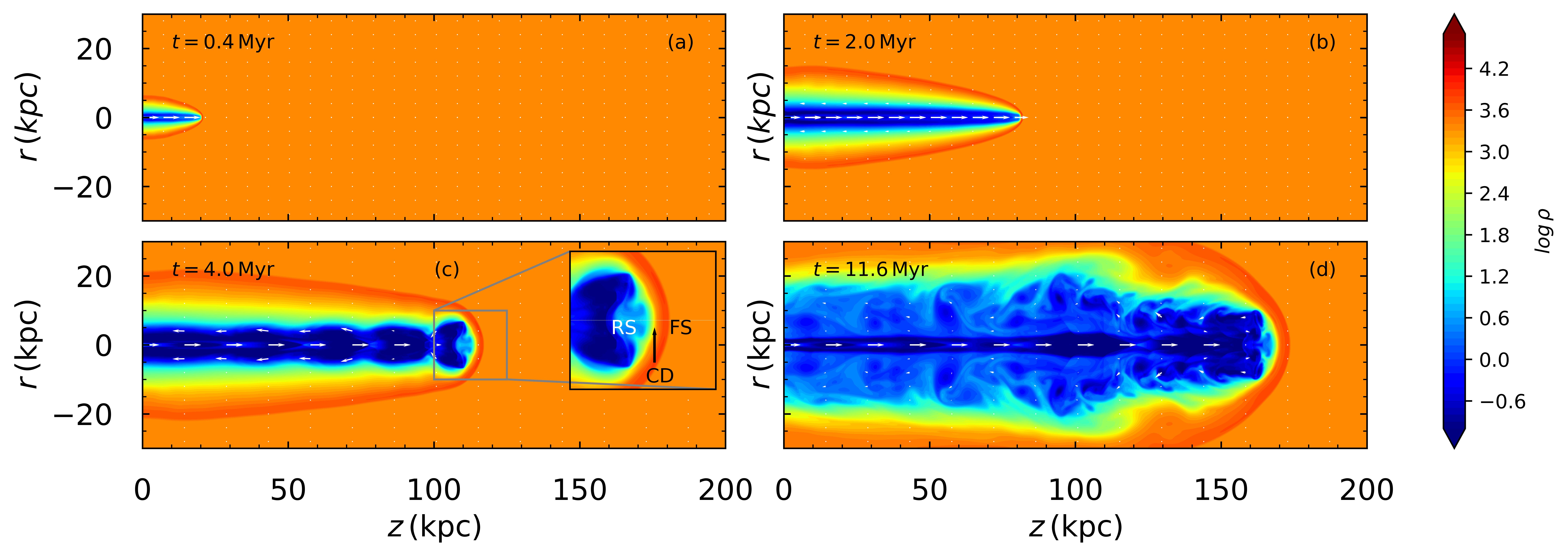} 
\caption{The density contours along with velocity vectors at various time steps as mentioned in the four panels (a --- $t=0.4$Myr, b --- $t=2.0$Myr,
c --- $t=4.0$Myr and d--- $t=11.6$Myr) for jet model OD. In panel (c) the FS, CD (jet-head) and RS are shown. The jet is composed of electrons and protons or $\xi=1$.}
\label{fig:OD1}
\end{figure}

In  Fig. \ref{fig:OD1} we have plotted the density contours along with the velocity vector field (white arrows) in the $r$-$z$ plane at different epochs to show the evolution of the jet of model OD. The supersonic jet advancing in a denser ambient medium drives a bow shock FS. The ambient medium material ahead of the jet head is pushed in the transverse direction, this leads to the formation of the overpressured and hot cocoon around the jet beam. In the initial phase of evolution, the jet remains thin and propagates as a very narrow beam of relativistic plasma. As the jet ploughs through the ambient medium, the jet head starts to expand and it develops backflow as shown in Fig.\ref{fig:OD1}(b). The zoomed inset in Fig.\ref{fig:OD1}(c) shows the structure of the jet head, and the location of FS, CD, and RS are shown in the inset. The sideways expansion and the formation of turbulent structures significantly reduce the propagation speed of the jet. Further in time $t\sim4$ Myr and $t\sim11.6$Myr, the interaction of the backflow and the jet beam intensifies, which causes the formation of multiple shocks, and a lot of structures start to form in the jet beam. In Fig.\ref{fig:OD1}(d) one can clearly see that the interaction of back-flowing jet material with the jet beam results in perturbations in fluids and these perturbations grow into the Kelvin-Helmholtz instability \citep{nwss82} which are transported back towards the jet nozzle as the jet advances ahead. The formation of vortex along the jet head and sideways expansion is clearly visible from the density contours (Fig.\ref{fig:OD1}d).

\begin{figure}[!h]
 \centerline {\includegraphics[width=15cm,height=6cm]
 {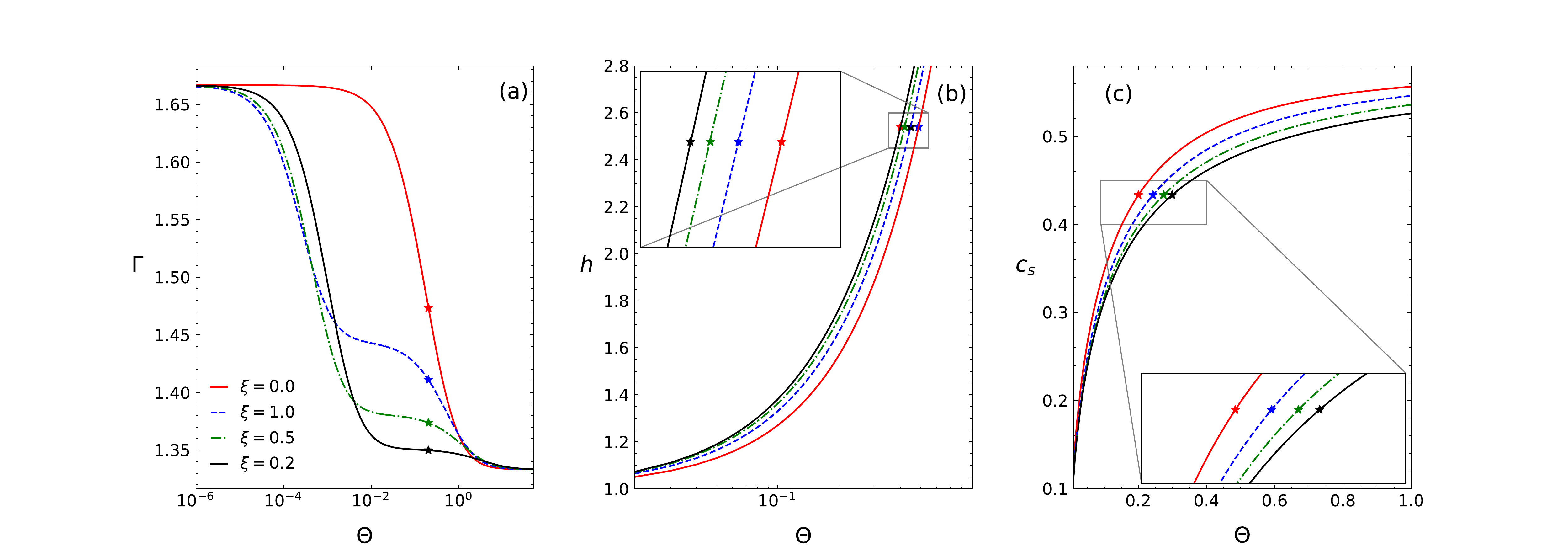}}
  \caption{The adiabatic index $\Gamma$ (a), enthalpy $h$ (b), and sound speed $c_s$ (c) as a function of $\Theta$ for various plasma compositions.}
  \label{fig:poly_vs_cs}
\end{figure}
  
\subsection{The thermodynamic variables}
It follows from eqs. \ref{eq:eos}---\ref{eq:poly}
any change in composition or $\Theta$ will affect
the thermodynamic quantities like the enthalpy, the sound speed, and the adiabatic index. In Fig. \ref{fig:poly_vs_cs}a-c we have plotted $\Gamma$, $h$, and $c_s$ as a function of $\Theta$, respectively. The composition parameter $\xi$ is mentioned in the figure. It is very clear that for $\Theta > 1$, $\Gamma \rightarrow~4/3$
and for low values of $\Theta$, $\Gamma \rightarrow 5/3$, for any values of $\xi$. However, for intermediate values, $\Gamma$ depends
on the composition of the gas. Since $\Gamma$ is the comparison of the random kinetic energy of the gas particles compared to the inertia of the same particles, its lower value indicates that the gas is thermally relativistic. At a given $\Theta$, the value of $\Gamma$ does not monotonically change with $\xi$. In some $\Theta$ range, $\xi=0.2$ is thermally the most relativistic, while at some other range $\xi=0.5$ is more relativistic. 
The stars on the curves in Fig. \ref{fig:poly_vs_cs}a, indicate the injection values of the jet in models A1-A4.
Similarly, $h$ is lowest for $\xi=0.0$ gas, while
for $\xi=0.2$ it is highest for most values of $\Theta$. So to attain the same value of $h$, the pair plasma needs to be very hot. Conversely, for a given value of $\Theta$, $c_s$ is highest for $\xi=0.2$. The stars in Fig. \ref{fig:poly_vs_cs}b
are the injection values for the jet in models B1-B4. And similarly, the stars in panel c are the injection parameters for models C1-C4.

\subsection{Model A: Same injection parameters} \label{sec:comp_effect}
\begin{figure}[!h]
 \centerline {\includegraphics[width=18cm,height=8cm]{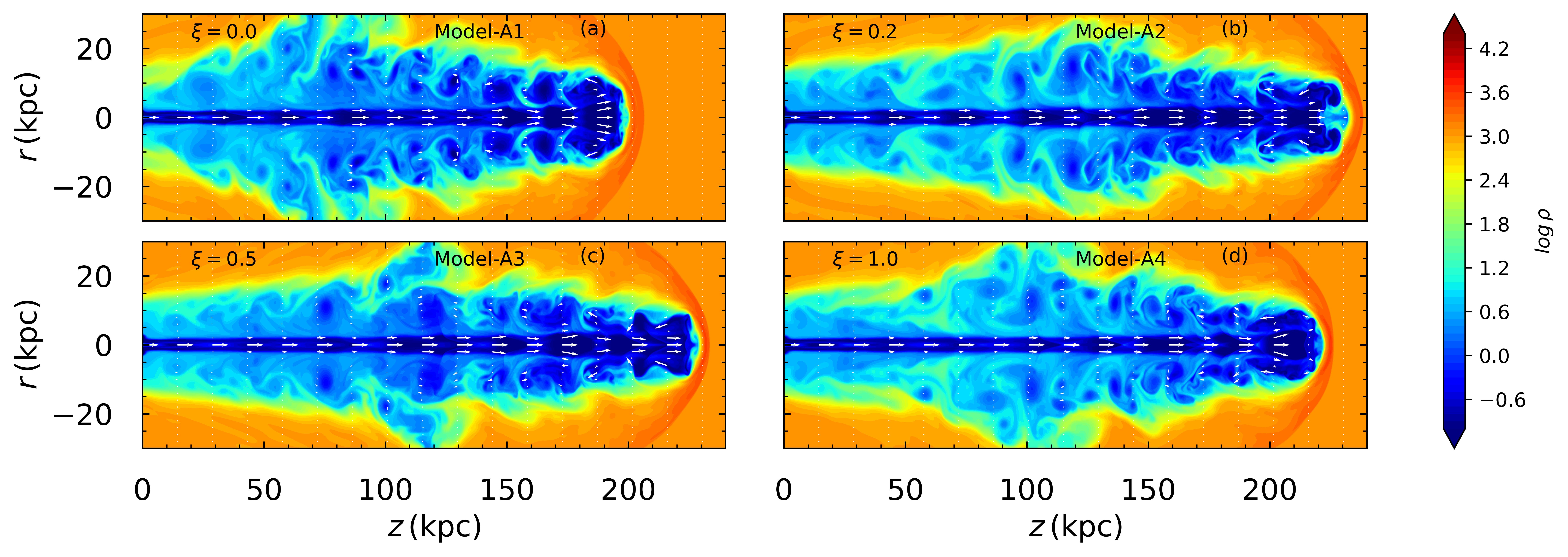}} 
  \caption{The contours of logarithmic density ($log\, \rho$) along with velocity vectors for various compositions. The injection parameters $\rhoj,\,\thetaj$ are kept same for all the jet models.}
  \label{fig:dens_same_inj}
\end{figure}

\begin{figure}[!h]
 \centerline {\includegraphics[width=18cm,height=8cm]{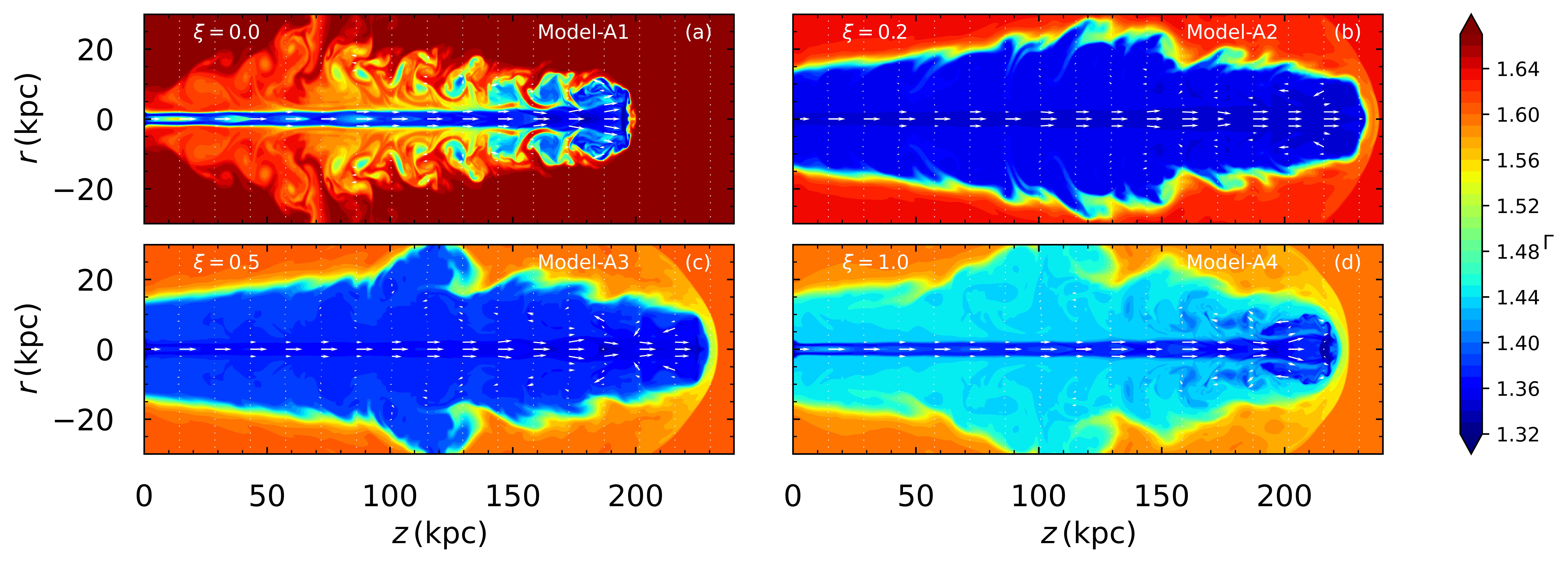}} 
  \caption{Contours of adiabatic index for different models (A1-A4) at $t=10.8 Myr$.}
  \label{fig:gamma_same_inj}
\end{figure}

The evolution of jet is characterized by the injection
velocity, the injected pressure, and density. In models A1--A4, we have kept the same injection values for velocity, pressure, and density $v_j$, $p_j$, and $\rho_j$ (see Table \ref{tab:model_param}). The difference between the models is in the compositions of the flow, in model A1 it is $\xi=0$, A2 it is $\xi=0.2$, in A3 $\xi=0.5$, and for A4 $\xi=1.0$.
 The stars in Fig. \ref{fig:poly_vs_cs}a show the values of the injection parameters which are used for the simulation of Models A1-A4. Despite fixing the value of $\Theta$ we can see that the various models have different sound speeds (hence different Mach numbers) and also a difference in the adiabatic index. 
In Fig. \ref{fig:dens_same_inj}a-d, we have plotted the density contours and velocity vectors of models A1-A4. All the panels are plotted at $t=10.8 Myr$. The model-A2 with composition $\xi=0.2$ is the fastest and with the highest jet kinetic power. Pair plasma or $\xi=0.0$ jet has the slowest propagation speed. The large scale vortices form at an earlier time for the pair plasma jet and hence it loses the thrust forward, resulting in a slower propagation speed. The jet structures behind the jet-head or CD are also different for different compositions. 
We have also plotted the corresponding contours of the adiabatic index ($\Gamma$) for models A1-A4 in Fig. \ref{fig:gamma_same_inj}a-d.
The beams of the jets with non-zero baryons i. e., (panels b, c, and d) are much hotter ($\Gamma \sim 4/3$), while the jet beam of pair-plasma jet (panel a) has regions where the jets become colder and the adiabatic index reaches values $\Gamma \sim 1.5$. 
The cocoon region is very hot for baryonic jets, but not so much
for pair-plasma jet. It was pointed out by \cite{cr09} that electron-positron plasma is thermally less relativistic (i. e., rest mass energy greater than the thermal energy) as is also shown in Fig. \ref{fig:poly_vs_cs}a. And since these simulations are for thermally driven jets, the electron-positron jet is slowest and less hot. This is also clear from the enthalpy column of Table \ref{tab:model_param}, where $h_j$ is greatest for the jet with $\xi=0.2$ (model A2), so the jet kinetic power ($\propto h$) will be highest for the jet of model A2 and lowest for model A1. It was also pointed out by \cite{cr09} and also from Table \ref{tab:model_param} of this paper, that electron-proton jet is not the most relativistic.

\subsection{Model B: Equal enthalpy models}

In model A, we launched jets with the same injection parameters, but different compositions of the flow changed the enthalpy of the jet. 
The specific enthalpy is related with the jet kinetic luminosity by
\begin{equation}
L_j=\gamma^2\rho h \pi R^2_b v_j \quad{erg\,s^{-1}}
\label{eq:jet_kl}    
\end{equation}
which can be written as
\begin{equation}
L_j=\gamma^2 \left(\frac{\rhoa}{\eta}\right) h \pi R^2_b v_j \quad{erg\,s^{-1}}; \qquad{\eta=\rhoa/\rhoj}
\label{eq:jet_kl_2}    
\end{equation}
Since the injection velocity is exactly same for all the models in this paper, therefore, from eq.
\ref{eq:jet_kl}, jet with higher values of $h_j$ (injected value of $h$) will have higher values of $L_j$ and therefore, it's no wonder that the jet morphology will depend on composition. So in model A, although the injection parameters are exactly the same, the composition is different so the jets corresponded to different jet power $L_j$. And it turned out that, the jet with $\xi=0.2$ has the highest $L_j$ for this value of $h_j$. It is expected that a jet with higher kinetic luminosity will produce higher propagation speeds and even the structures formed are influenced by $L_j$ \citep{shr21,rbmc20,mbrm16}. Hence, in order to check the effect of composition, we tune the value of $\Theta_j$ in such a way that all the models with different compositions have the same specific enthalpy and therefore with the same jet power. Assuming the ambient medium density of the order of $10^{-4}{\rm m_p/cm^3}$ \citep{f98}, we have calculated the jet power for all the models, which is quoted in Table \ref{tab:model_param}.

\begin{figure}[!h]
 \centerline {\includegraphics[width=18cm,height=8cm]{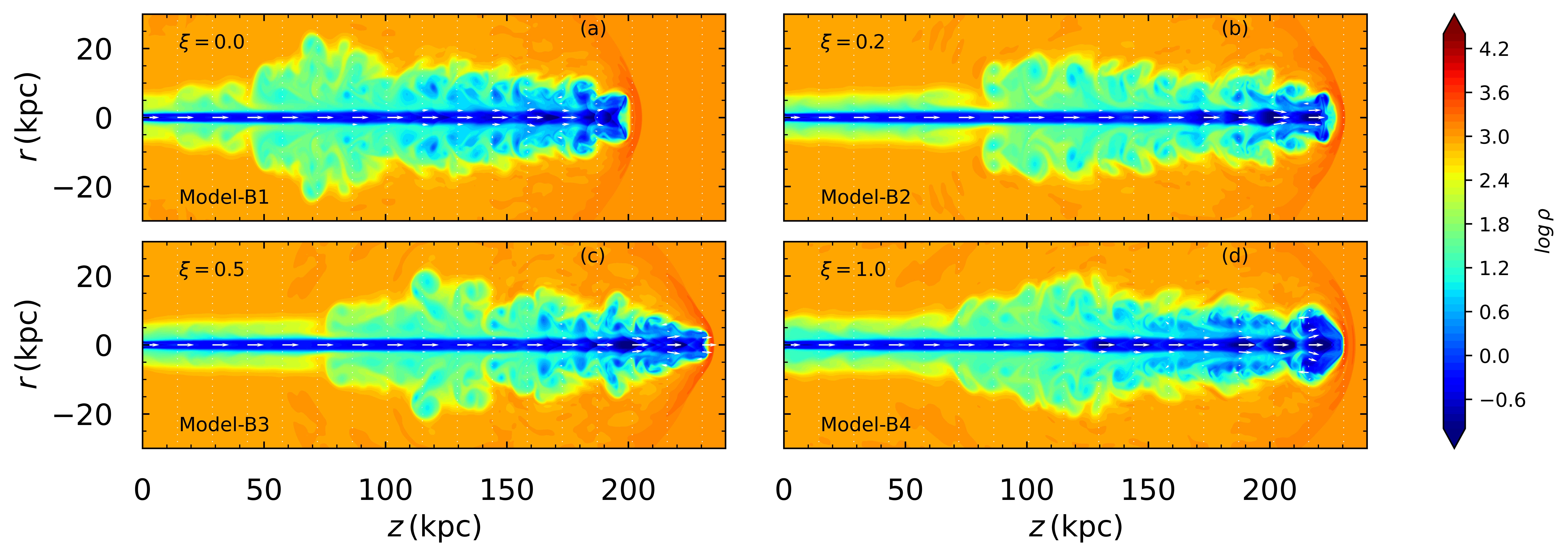}} 
  \caption{The density contours along with velocity vectors for various compositions. The specific enthalpy is kept same for all the jet models B1-B4.}
  \label{fig:same_enth}
\end{figure}

\begin{figure}[!h]
 \centerline {\includegraphics[width=18cm,height=8cm]{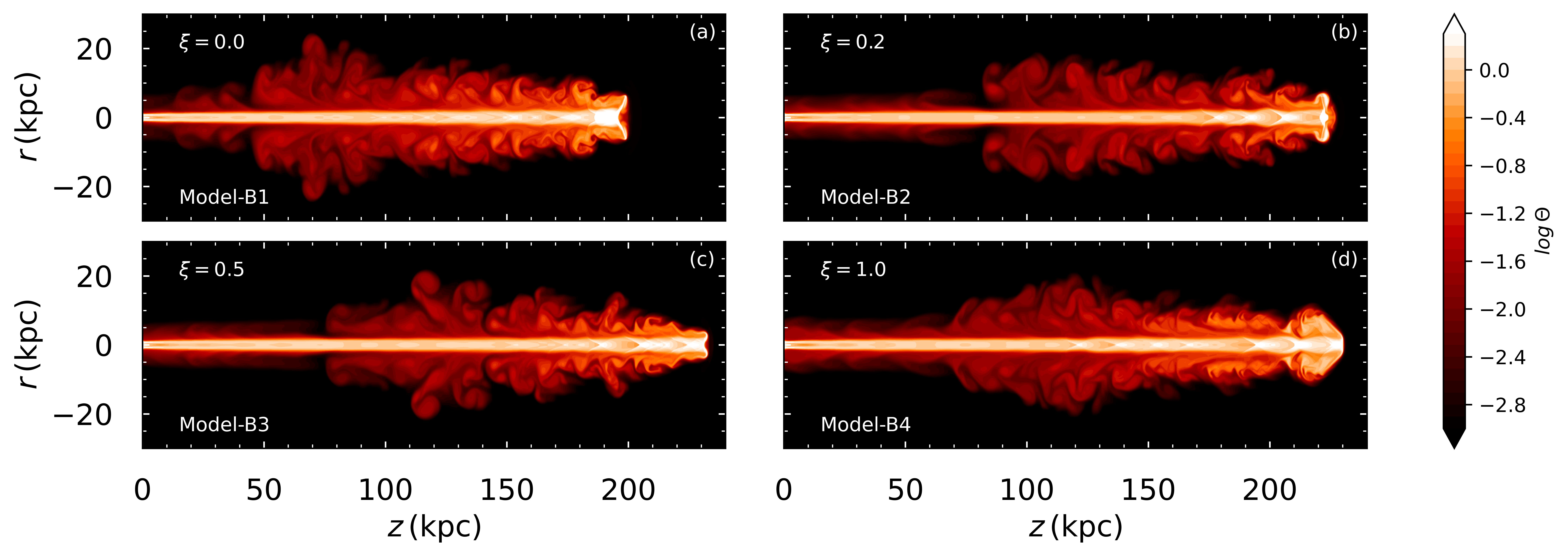}} 
  \caption{The temperature ($\Theta$) contours for models B1-B4.}
  \label{fig:temp_same_enth}
\end{figure}

In Fig. \ref{fig:same_enth}(a)-(d) density contours and velocity vectors are plotted at $t=8.0\,{\rm Myr}$ for the jet models B1-B4. The injection parameters correspond to the stars in Fig. \ref{fig:poly_vs_cs}b. We show that the propagation speeds and the structures formed by the jets even after fixing the jet kinetic power differ with change in composition. In comparison to other plasma compositions, $\xi=0.0$ shows a higher number of turbulent structures. The jet head area is narrower for composition $\xi=0.5$ in comparison to other models. Once again the propagation speed of the jet with $\xi=0.0$ is the slowest, however, in this particular case $\xi=0.5$ jet is fastest, followed by $\xi=1.0$. Interestingly, out of the four different models of jet with the same enthalpy, $\xi=0.5$ has the strongest recollimation shock (compression ratio $=41$ measured along the axis), while $\xi=0.2$ has the strongest reverse shock (compression ratio $=16.7$, measured along the axis). In Fig. \ref{fig:temp_same_enth} a-d, we have plotted the contours of $\Theta$. The $\Theta$ distribution shows the locations of hot-spots for these models. The temperature contours show that the locations as well as the areas of the hot-spots are different for these models. The structure of the Mach disk changes for different compositions.

\begin{figure}[!h]
 \centerline {\includegraphics[width=18cm,height=8cm]{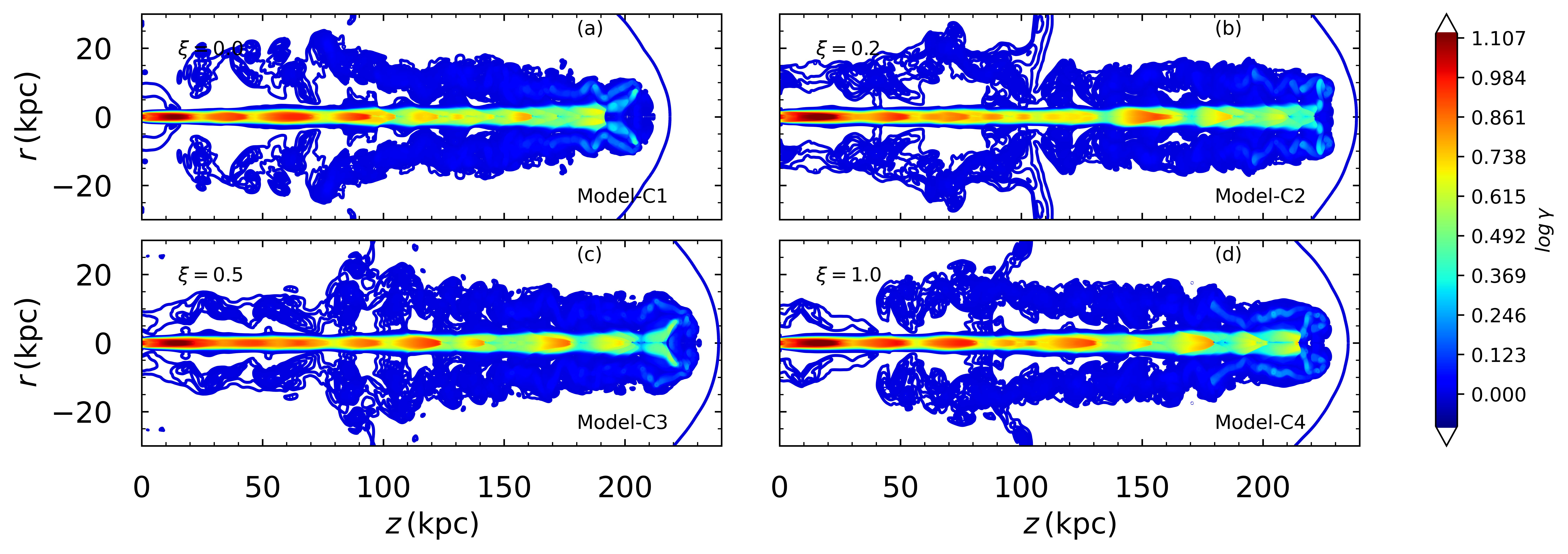}} 
  \caption{The contours of Lorentz factor for different models C1-C4.}
  \label{fig:same_mach}
\end{figure}

\begin{figure}[!h]
 \centerline {\includegraphics[width=14cm,height=14cm]{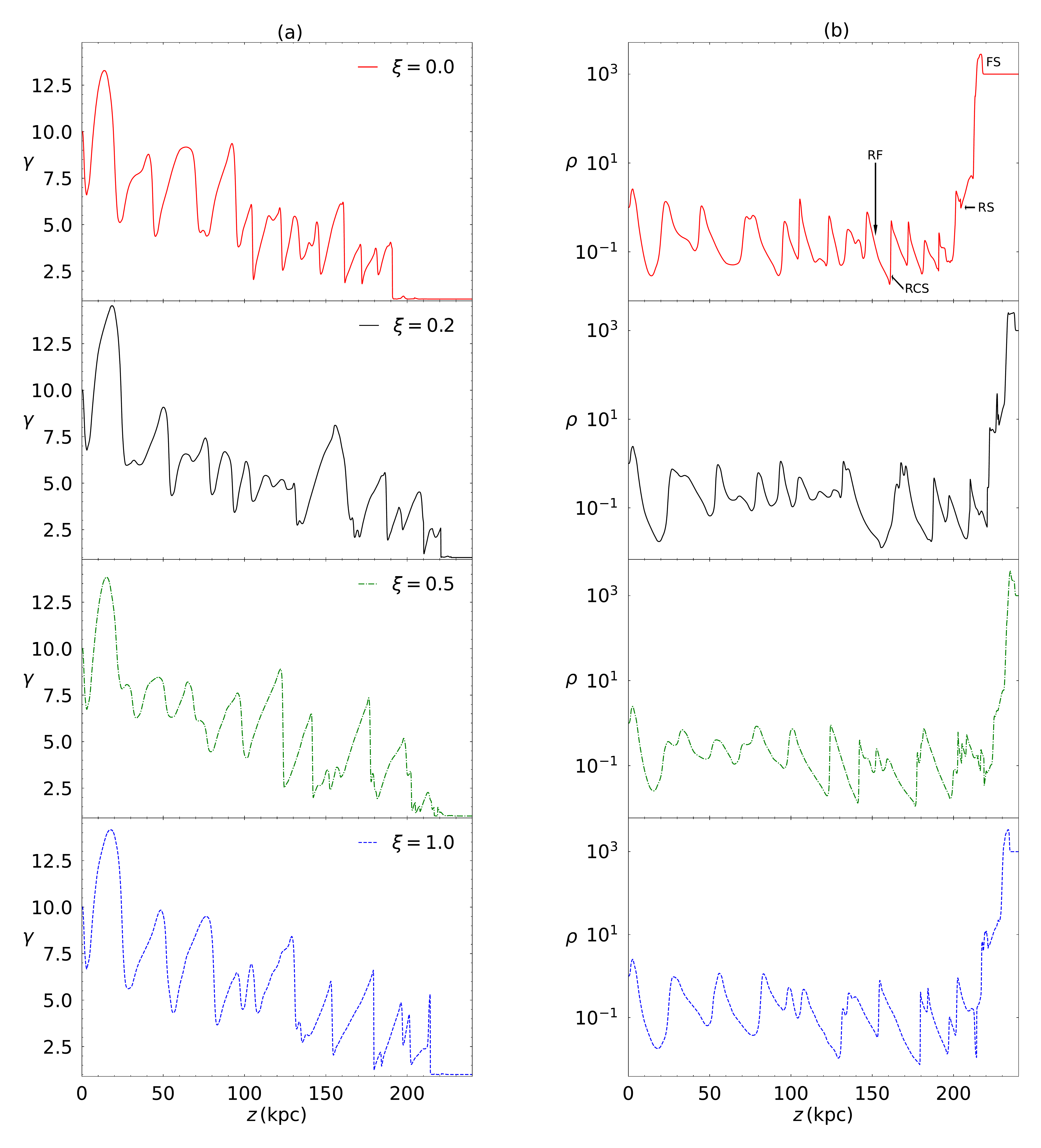}} 
  \caption{(a) Jet Lorentz factor $\gamma$ and corresponding (b) $\rho$ as a function of $z$ for different jet compositions at $t=11.6\,Myr$. FS, RS and the recollimation shock (RCS) are marked in the top most panel of column b. The flow variables along the spine of the jet corresponds to models C1-C4.}
  \label{fig:1d_dens_same_mach}
\end{figure}

\subsection{Model-C: Same Mach number jets}
The structure of the cocoon crucially depends on the Mach number of the jet. So, it is expected that the structure of the jet head and cocoon will be different if the Mach number of the jet beam changes, and in model B the Mach number of the jets were different although the jet power was same. Now we investigate the case when the injected value of the Mach number of the jet is kept same for different values of composition. The value of $\Theta_j$ for Model C1-C4 is tuned in a way that the injected value of the sound speed of the jet beam remains same in every model even after the variation in the composition parameter. Since all the jets are also launched with the same velocity, therefore in this case, it implies that they are also launched with the same Mach number. The value of $\Theta_a$ is also tuned such that the value of local sound speed in the ambient medium is same for all the models.
The injection parameters of models C1-C4 correspond to the stars in Fig. \ref{fig:poly_vs_cs}c. The contours of $\log\,\gamma$ at $t=11.6\,Myr$ are plotted in Fig. \ref{fig:same_mach}a-d for these simulations. The locations of recollimation shock in the jet beam are most clearly visible in this figure. The FS is the leading blue curve. The jet head or CD has a dimple for the $\xi=1.0$ jet (model C4). The propagation speed of the $\xi=0.0$ jet is the slowest while those due to $\xi=0.5$ and $\xi=0.2$ are similar. 

\begin{figure}[!h]
 \centerline {\includegraphics[width=14cm,height=8cm]{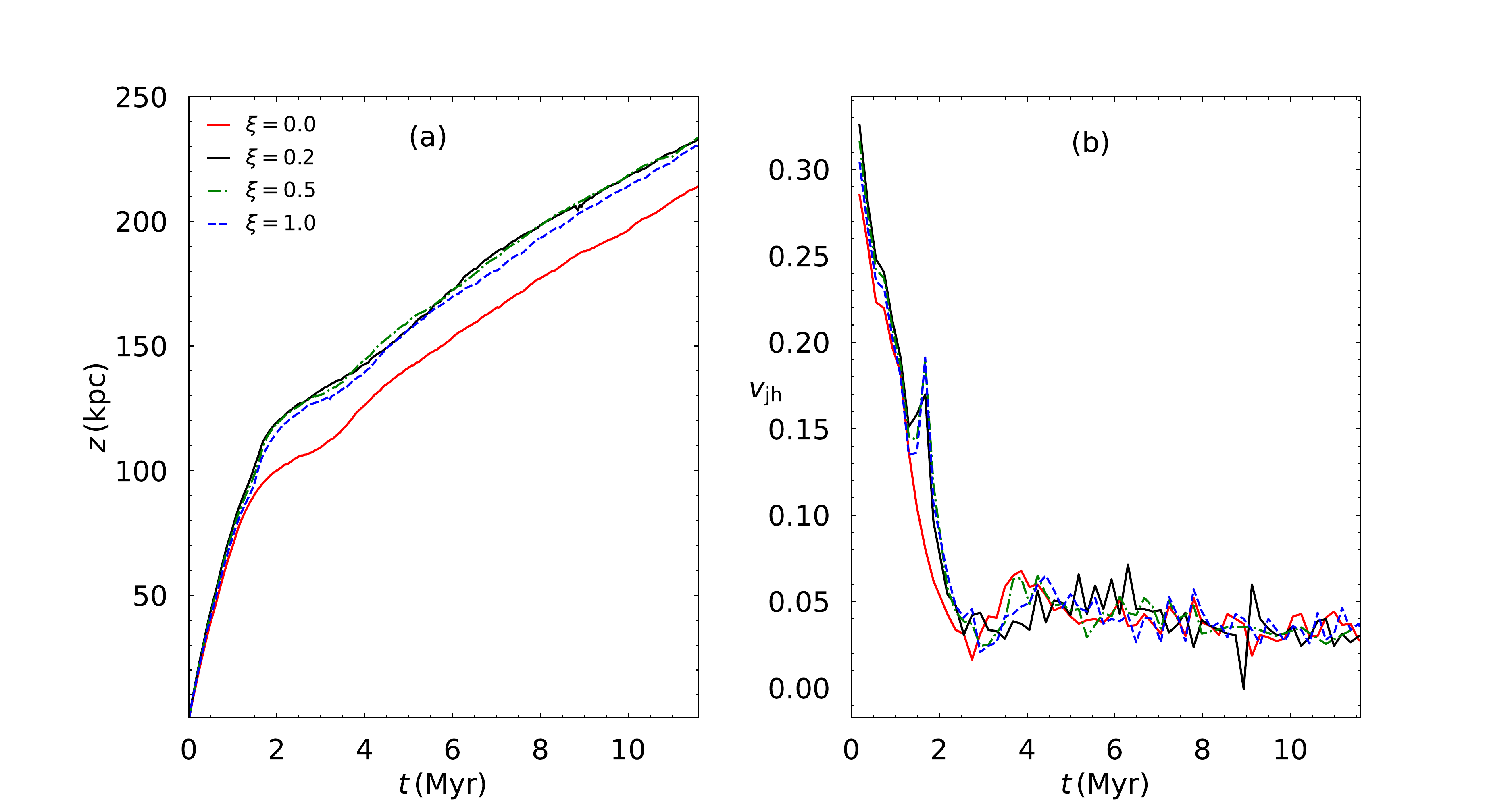}} 
  \caption{(a) Jet head position and (b) jet-head velocity $v_{jh}$ as a function of $t$ for different jet compositions. The compositions marked in panel a, corresponds to models C1-C4.}
  \label{fig:pos_same_mach}
\end{figure}

It is intuitive that the jet will lose speed as it wades through the ambient medium due to the resistance offered by the medium. However, a large number of numerical simulations have shown that for some initial conditions, jets can be accelerated above the injected Lorentz factor by a significant margin. One of the simplest but very strong hydrodynamic booster was suggested by \cite{ar06}. The production of a strong rarefaction wave near the contact discontinuity accelerates the jet with an increase in overall propagation speed \citep{ajm05,hmzbcss17}. In addition, the
formation of many recollimation shocks (RCS) within the jet beam, may locally decelerate the flow, but the associated rarefaction fan (RF) may enhance the local speed of the jet. In order to highlight this, we have plotted the local Lorentz factor along z axis i.e., $\gamma(0,z)$ (Fig. \ref{fig:1d_dens_same_mach} a) and density $\rho~(0,z)$ (Fig. \ref{fig:1d_dens_same_mach}b) as a function of $z$. 
The panels (top-bottom) in columns (a) and (b) of Fig. \ref{fig:1d_dens_same_mach}, show the variation of Lorentz factor and density for models C1 ($\xi=0$)-C4 ($\xi=1.0$), respectively. In the top panel of \ref{fig:1d_dens_same_mach}(b) we have marked the positions of FS, RS, RCS, and RF. The Lorentz factor is boosted to a value higher than the injected value in the first RF for all the models.      
Likewise, all the sharp jumps in density and Lorentz factor are the locations of internal shocks in the beam. The rarefaction fans are the regions between two internal shocks as shown in Fig. \ref{fig:1d_dens_same_mach}b. The rarefaction wave is an expansion wave hence the velocity/Lorentz factor increases in this region, which is clearly visible. 
So there would be alternate denser and rarer regions and correspondingly slower and faster regions within the jet beam. Therefore, if an RCS and RF combination forms near the jet head then the jet propagation speed should also be higher. In Fig. \ref{fig:pos_same_mach}a,
we plot the jet head positions on the jet axis as a function of time for the four jets described in models C1-C4. And in Fig. \ref{fig:pos_same_mach}a,
we plot the corresponding jet head velocity $v_{jh}$ as a function of time. From the jet head positions, it is clear that the $\xi=0.0$ jet (model C1) is the slowest.
And out of the four jets, $\xi=1.0$ (model C4, blue dashed) is slower than the other two jet types. 
For these injection parameters, the jets decelerate at some time (around $2$ Myr for $\xi=0.0$ jet) but re-accelerates (around $4$ Myr for $\xi=0.0$). These are due to the formation of a combination of RF-RCS near the jet head. For the jets containing baryons this reacceleration is somewhat at an earlier time. This deceleration and re-acceleration
of the jets are also clear from Fig. \ref{fig:pos_same_mach}b, where we plot the velocity of the jet head surface along the jet axis. Although overall the propagation speed of the pair-plasma jet is lower but can overshoot the speed of the jet-head for other baryon containing jet.

\section{Discussion and Conclusions}
\label{sec:disc}
\subsection{Discussion}

The morphology of thermally driven relativistic jets has been studied by various authors
\citep{mmi97,samgm02,myt04} each author focusing on large scale evolution and dynamics of relativistic jets. In this paper, we have investigated the propagation of relativistic jets through a uniform ambient medium, using an approximate but analytical relativistic equation of state (CR EoS), by performing axisymmetric, relativistic, hydrodynamic simulation in cylindrical geometry. 
The code used in this paper is based on Harten's TVD routine but adopted for relativistic hydrodynamics and relativistic EoS \citep{rcc06,crj13}. 
Since relativistic flows are hot, ionized, and can be thermally transrelativistic, the fixed $\Gamma$ EoS is not compatible with the physics of the flow. A relativistic plasma is composed of electrons and ions, and the information of the constituent particles is in the equation of state. The equation of state on the other hand appears in the expression of state variables like momentum density and energy density of the flow. Therefore, the evolution of the flow should depend, in addition to other things, on the composition too. The effect of composition has been shown in the case of 
accretion onto compact objects \citep{cr09,cc11, ck16,scp20}. In order to reduce the number of free parameters we fixed the injection velocity $v_j=0.995$ fixed for all the models. We also fixed the jet cross-section and the ambient medium density for all the models. Initially, we ran a generic jet simulation for electron-proton ($\xi=1.0$) jet (model OD) and showcase the generic features like the FS, CD, and RS, the back flowing jet material, in addition to the formation of various recollimation shocks along the beam of the jet. Then we considered models A1-A4 all starting with the same $\Theta_j$ but took four different values of $\xi$. In Fig. \ref{fig:poly_vs_cs}a injection parameters are the stars on the curve, the adiabatic indices $\Gamma_j$ and the corresponding sound speeds are different. This means that the eigenvalues of the flow will differ for jets with different compositions, and therefore the eventual evolution of the jets with different compositions would be different in terms of propagation speed, and also the structure of the jet  as is shown in Figs. (\ref{fig:dens_same_inj}, \ref{fig:gamma_same_inj}). The adiabatic index or $\Gamma$ contour plots also show that it is the interaction that determines the value of the adiabatic index along the jet beam and also in the  cocoon. Electron-positron jets also shows a lot of structure and are generally colder ($\Gamma$ values are higher). In models B1-B4, jets with the same enthalpy are launched (stars in Fig. \ref{fig:poly_vs_cs}b), and since all start with the same injection velocity, therefore, it means all of these jets are launched with the same jet kinetic power. Even in this case, the jet propagation velocity and detailed structures depend on the composition parameter $\xi$. The jet with $\xi=0.5$ is the fastest and $\xi=0.0$ is the slowest, and turbulent structures are more pronounced compared to jets containing protons. In the recollimation shock regions, the density enhanced regions appear to coincide with the hot regions but not so in the jet head region. This would have an interesting effect on the brightness and spectral morphology of the jets. The number of collimated shock regions is also more compared to $\xi=0.0$ jet. In models C1-C4, jets launched with the same Mach number, but different compositions are investigated. The Lorentz factor contours exhibited the recollimation shocks more vividly. With these injection values, the $\xi=0.2$ and the $\xi=0.5$ attain comparable speed, although the structures are different. The jet head structures are different as well, the electron-proton jet head shows an inward dimple. In fact between $t=2-4$Myr, $\xi=0.0$ jet head speed exceeds that due to all other proton containing jets. But the initial inertia of proton carrying jets keeps those jets ahead of the pair plasma jet.
We have compared jets with the same injection parameters, same enthalpy and injection speed and same Mach number and injection speed, but with different compositions. Since the information of composition is in the local enthalpy of the jet, therefore the evolution of the jet in terms of propagation speed, or the number of recollimation shocks, or the structure of the Mach disk etc depends on the composition, but the basic structure of jet with a forward shock, jet-head (or contact discontinuity) and a reverse shock is seen for jets of all compositions. Additionally, we have performed two dimensional (2D) axisymmetric simulation of jets, but the recent three dimensional (3D) simulations \citep{rbmc20,shr21} highlight that differences in dimensionality can lead to different physical outcomes. One may study the non-axisymmetric interaction of jets with the ambient medium. It may produce more complex structures. In 2D energy cascade to smaller scales is not possible, the properties of turbulence in 3D simulations are radically different \citep{mbrm16}. Hence, 3D simulations can provide more realistic and complex results to quantify the effect of composition on the dynamics and structures of the jets.

\subsection{Conclusions}
Numerical simulation of the relativistic jet is a time-dependent study of a supersonic, relativistic beam of matter flowing through a denser, colder medium.
From a large number of simulations, it is common knowledge that jets launched with the same injection parameters should evolve identically. However, our results (Models A1-A4) show that despite fixing the injection parameters, jets with different compositions show a difference in terms of propagation speeds and also the structures formed by the jets. The electron-positron jets are less relativistic (less hot and slower) than jets with all other composition parameters. We also studied jet models (B1-B4) with same kinetic luminosity but with different plasma compositions. Even after fixing the jet kinetic luminosity, we found out that the propagation velocities are different, and the jet with $\xi=0.5$ 
turned out to be the fastest and the jet composed of pair plasma is slowest. The boosting mechanisms are important aspects of multi-dimensional jet simulations and the jets launched with the same initial Mach number (Models C1-C4) show that the reacceleration epochs vary with change in the composition parameter. The electron-positron jets can overshoot the speed of the jet-head for other baryon containing jets, but overall it remains slower than the jets containing baryons. It may be noted that we have chosen initial flow parameters for different models which did not cause widely different injection values ($L_j$, $h_j$ etc are close by) yet there is a notable difference between the pure leptonic and jets with baryon content. Conclusions might have been different if we had chosen other parameters. For example, in Fig. \ref{fig:poly_vs_cs} a, the $\Theta_j$
chosen keeps the $\xi=0.2$ jet as thermally most relativistic, but at lower $\Theta_j$ flows with other $\xi$ might be more relativistic. All these cases show that the composition of the jet is important in the eventual propagation and morphology of the jet and it cannot be predicted or parameterized apriori.

\begin{acknowledgments}
We would like to thank the anonymous referee for their valuable comments and suggestions that greatly improved the quality of this manuscript.
\end{acknowledgments}

\bibliography{biblio}{}
\bibliographystyle{aasjournal}



\end{document}